\documentclass[twoside,a4paper,11pt]{sea10}
\usepackage{graphicx}
\usepackage{hyperref}
\usepackage{movie15}
\begin{document}
\pagenumbering{arabic}
\pagestyle{myheadings}
\thispagestyle{empty}
{\flushleft\includegraphics[width=\textwidth,bb=58 650 590 680]{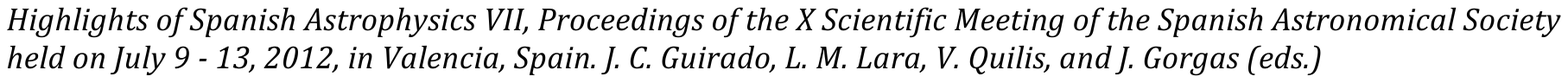}}
\vspace*{0.2cm}
\begin{flushleft}
{\bf {\LARGE
%
Collaborating with ``professional'' amateurs:
low-mass stars in fragile multiple systems 
%
}\\
\vspace*{1cm}
%
J. A. Caballero$^{1}$,
J. Genebriera$^{2}$, 
T. Tobal$^{3}$, 
F. X. Miret$^{3}$, 
F. M. Rica$^{4}$, 
J.~Cairol$^{3}$, 
N. Miret$^{3}$, 
I. Novalbos$^{3}$, 
D. Montes$^{5}$ 
and 
A. Klutsch$^{5}$
%
}\\
\vspace*{0.5cm}
%
$^{1}$ Centro de Astrobiolog\'{\i}a (CSIC-INTA), European Space Astronomy Centre, PO Box 78, 28691 Villanueva de la Ca\~nada, Madrid, {\tt caballero@cab.inta-csic.es}\\
$^{2}$ Observatorio de Tacande, La Palma\\
$^{3}$ Observat\`ori Astronomic del Garraf, Barcelona, {\tt www.oagarraf.net}\\
$^{4}$ Agrupaci\'on Astron\'omica de M\'erida, Badajoz\\
$^{5}$ Departamento de Astrof\'{\i}sica y Ciencias de la Atm\'osfera, Facultad de Ciencias F\'{\i}sicas, Universidad Complutense de Madrid, 28040 Madrid
%
\end{flushleft}
%
\markboth{
``Professional'' amateurs and fragile multiple systems
}{ 
%
Caballero et al. 
%
}
\thispagestyle{empty}
\vspace*{0.4cm}
\begin{minipage}[l]{0.09\textwidth}
\ 
\end{minipage}
\begin{minipage}[r]{0.9\textwidth}
\vspace{1cm}
\section*{Abstract}{\small
%
The boundary between professional and amateur astronomers gets narrower and narrower. 
We present several real examples, most of them published in refereed journals, of works resulting from fruitful collaborations between key amateur astronomers in Spain and professional colleagues. 
The common denominator of these works is the search for binaries, mostly nearby, wide, common proper-motion pairs with low-mass stellar components, including some of the most fragile systems ever found.
%
\normalsize}
\end{minipage}
%
%
%
\section{Today: {\em in media res} \label{section.1}}

In the 17th century some good practices for scientific writing were already established, and since then scientists have given more importance to plain and accurate description than rhetorical flourishes.
However, authors in that epoch also emphasised the importance of not boring the reader.
Here we present the results of several professional-amateur collaborations on low-mass stars in fragile multiple systems using {\em in media res} (``into the midst of things''), the literary and artistic narrative technique wherein the relation of a story begins either at the midpoint or at the conclusion, rather than at the beginning.
We wish the reader to feel like enjoying a bloodless Quentin Tarantino's film script: into the middle of a sequence of events that led a group of amateur astronomers make great discoveries with only sporadic supervision from professionals.

\section{\label{section.200807} July 2008: proper-motion companions to Luyten stars}

\begin{figure}
\center
\includegraphics[width=0.75\textwidth]{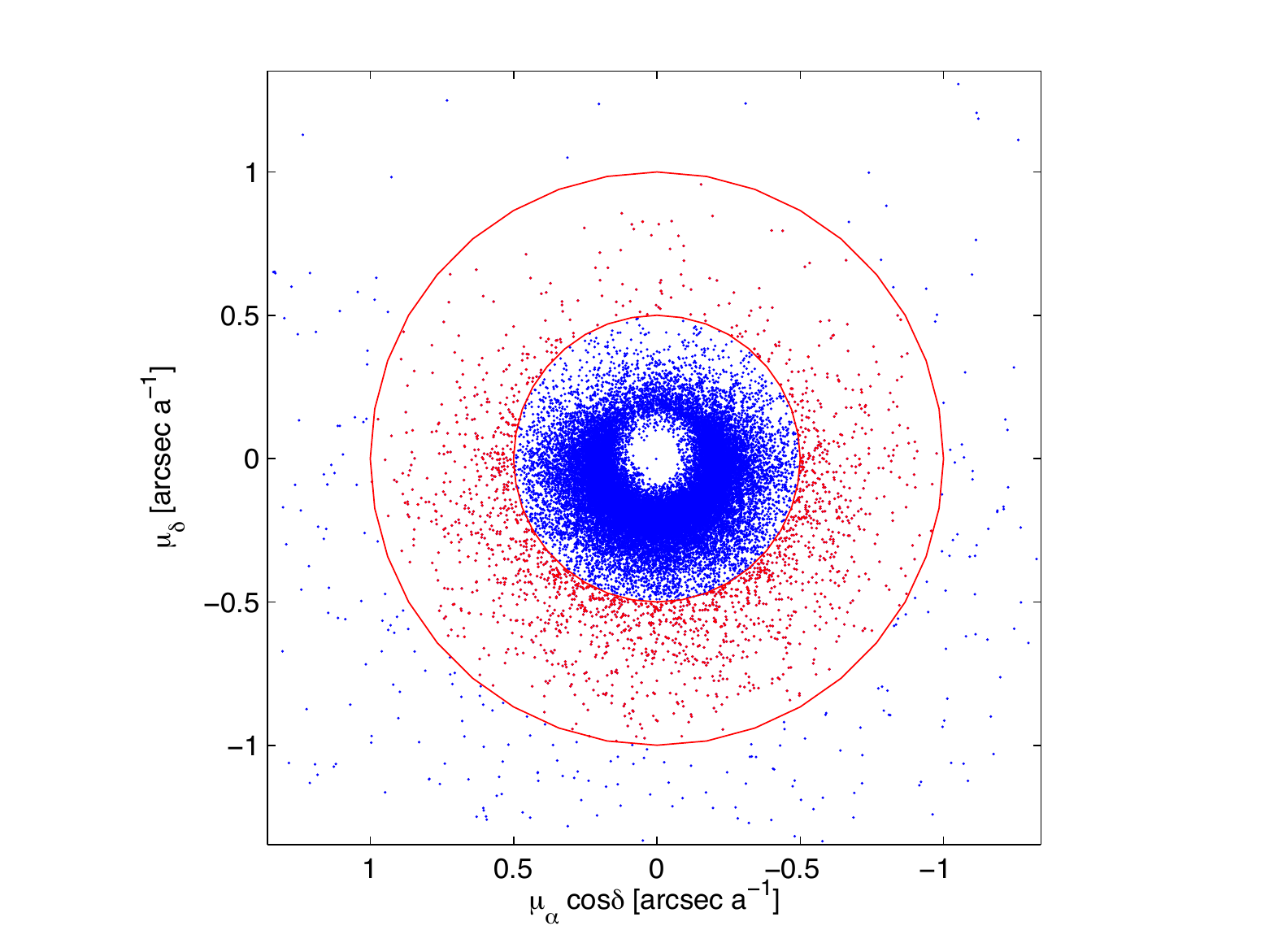} 
\caption{\label{F1} Inner part of the proper-motion diagram ($\mu_\alpha \cos{\delta}$ vs. $\mu_\delta$) of the Luyten stars in Salim \& Gould (2003).
The [red] corona indicates the interval of proper motions studied by Caballero et~al. (2010a).}
\end{figure}

This contribution corresponds to a talk given by the first author during the outreach session of the X meeting of the Spanish Astronomical Society (Sociedad Espa\~nola de Astronom\'{\i}a, SEA) in Valencia in summer 2012.
Four years before, at the VIII SEA meeting in Santander, he gave two talks: one on formation, evolution and multiplicity of brown dwarfs and giant exoplanets (Caballero 2010a), the other on a Virtual Observatory search for companions to Luyten stars in close collaboration with some of the authors of this proceeding.
There, Caballero et~al. (2010a) showed preliminary results of an Aladin-based survey for companions to high-proper-motion Luyten stars in the Salim \& Gould (2003) catalogue.   
Among the 1947 studied stars (and white dwarfs) with proper motions in the interval 0.5 $< \mu <$ 1.0\,arcsec\,a$^{-1}$ (see Fig.~\ref{F1}), they recovered 76 double, triple and quadruple systems, many of which are still poorly known.
Measured angular separations ranged from about 5 to over 500\,arcsec.

\begin{table}[] 
\caption{Some investigated systems with Luyten primaries wider than $\rho$ = 100\,arcsec} 
\center
\begin{minipage}{1.0\textwidth}
\center
\begin{tabular}{l llc cc} 
\hline\hline 
WDS\footnote{Washington Double Star catalogue.} & Primary 		& Secondary	& Sp.		& $\rho$ 		& $\theta$	\\  
code	 		& 	 		& 	 		& types					& [arcsec]		& [deg]	\\ [0.5ex] 
\hline  
STT 547 AF	& HD 38 AB  	& BD+44 4548	& (K6\,V+M0\,V) + M1\,Ve		& 327.9		& 253.8	\\ 
LDS 71		& L 512--14 	& L 512--15 	& M3\,Ve + M:				& 105.4		& 312.1	\\ 
...\footnote{The system AB--C, discovered by Caldwell et al. (1984), has no WDS entry yet.}	& HD 15468 AB & HD 15468 C	 & (K4.5\,V+K:\,V) + M2.5\,V		& 473		& 242.1	\\ 
LDS 3508		& CD--31 1454 & LP 888--25 	& K2:\,V + K:\,V				& 223.0		& 263.5	\\ 
RUB 4 		& BD--17 3088 	& WT 1759 	& K7\,V + DZ\,7				& 399.0		& 192.8	\\ 
LDS 4172		& LP 794--17 	& LP 794--16 	& K: + M:					& 323.7		& 293.7	\\ [0.5ex] 
\hline
\end{tabular} 
\end{minipage}
\label{T1} 
\end{table}

Six of the widest systems identified by Caballero et~al. (2010a), with angular separations greater than 100\,arcsec, are listed in Table~\ref{T1}.
Only three secondaries, two early-M dwarfs and one white dwarf, have been spectroscopically studied in detail.
The other three secondaries have only rough spectral-type estimations and deserve a further astrometric, photometric, and spectroscopic analysis.
Caballero et~al. (2010a) also serendipitously re-discovered half a dozen poorly-known physical pairs, one of which outstood because of its supposed youth, late spectral type, nearness and wide projected physical separation: GIC~158. 


\section{\label{section.201007} July 2010: G~125--15\,AB + G~125--14 = GIC 158}

Caballero et~al. (2010b) investigated in detail the system GIC~158, whose primary (G~125--15\,AB) is an active M4.5\,Ve star previously inferred to be young ($\tau \sim$ 300--500\,Ma) based on its high X-ray luminosity.
Actually, it is an inflated, double-lined, spectroscopic binary with a short period of photometric variability of 1.6 days.
The observed X-ray and H$\alpha$ emissions, photometric variability, and abnormal radius and effective temperature of the primary are indicative of strong magnetic activity, possibly because of orbital synchronisation and rapid rotation.
The secondary (G~125--14) has the same spectral type but is more than one magnitude fainter than the primary.
At $d \sim$ 26\,pc, the estimated projected physical separation between the two components of about 1.2\,kAU ensures that GIC~158 is one of the widest systems with intermediate M-type primaries yet found in the solar neighbourhood.

\section{\label{section.200700} From November 2006 to March 2007: Koenigstuhl 1,~2,~3}

Years earlier, while preparing a near-infrared colour-colour diagram of young brown dwarfs, Caballero (2007a) serendipitously discovered the widest ``ultracool'' binary at that time. 
Koenigstuhl~1 is a common proper-motion pair of M6.0:\,V and M9.5\,V stars separated by 78\,arcsec. 
At $d \sim$ 23\,pc, the projected physical separation becomes 1.80$\pm$0.17\,kAU, which was three orders of magnitude greater than the widest separation between ``normal'' ultracool binaries then known.
Because of its low total mass, of only about 0.18\,$M_\odot$, its binding energy makes the pair be still one of the most fragile systems known to date.

Caballero (2007b) went on looking for common proper-motion companions to almost 200 stars and brown dwarfs with spectral types later than M5.0\,V in the southern hemisphere, and discovered Koenigstuhl~2 and 3. 
The latter is a multiple system formed by a bright, slightly metal-poor, F8\,V star, an M8.0\,V+L3\,V close pair at the astonishing projected physical separation (for its mass) of 11.9$\pm$0.3\,kAU, and another L1\,V close companion to the primary recently discovered by Gauza et~al. (2012).

\section{From May 2009 to December 2009: the widest double stars}

Our Aladin-based proper-motion survey of Luyten stars (Section~\ref{section.200807}), which actually was a heritage of the Koenigstuhl survey (Section~\ref{section.200700}), resulted not only in the characterisation of the magnetically-active, low-mass, triple system GIC~158 (Section~\ref{section.201007}), but also in the recovery of fragile pairs with angular separations greater than 100\,arcsec and total masses less than 1\,M$_\odot$. 
But are they actually bound?

Before going ahead with our study, we had to make sure of that some of our discoveries had a physical meaning. 
There were already very wide binaries and candidates in the literature with projected physical separations $s >$ 50\,kAU (e.g., Gliese 1969; Allen et~al. 2000; Zapatero Osorio \& Mart\'{\i}n 2004; L\'epine \& Bongiorno 2007; Makarov et~al. 2008; Poveda et~al. 2009; see also Scholz et~al. 2008).
However, numerous works had concluded that dynamical evolution dictates a sharp cutoff in the number of very wide binaries with physical separations greater than about 20\,kAU (one tenth of a parsec -- e.g., Tolbert 1964; Bahcall \& Soneira 1981; Abt 1988; Wasserman \& Weinberg 1991; Poveda \& Allen 2004).

In his first issue of the series ``Reaching the boundary between stellar kinematic groups and very wide binaries'' at Astronomy \& Astrophysics, Caballero (2009) concluded that the key parameter for ascertaining the physical meaning of an ultrawide pair ($s >$ 20\,kAU) is not the projected physical separation itself, but the gravitational binding energy, defined by $U_g^* = -G M_1 M_2 / s$.
Besides, he found that the most fragile systems, regardless of their total mass, have binding energy moduli of the order of 10$^{33}$--10$^{34}$\,J.

Caballero (2010b) took the statement ``size {\em and mass} matter'' to the limit with the identification of the star KU~Lib as the distant fifth member of the Zubenelgenubi ($\alpha$~Lib) multiple system based on common proper motion, parallax, radial velocity and age.
The stars, which are part of the young Castor moving group ($\tau \sim$ 200\,Ma), may be in the process of disruption.
With a total mass of 6.7\,M$_\odot$, in spite of the gargantuan projected physical separation of over 200\,kAU (about 2.6\,deg projected in the sky), the binding energy modulus is not lower than, for example, the nearby system $\alpha$~Cen~AB + Proxima Centauri.
The confirmation that this kind of ultrawide systems does exist was afterwards brought by Shaya \& Olling (2011) and Makarov (2012).
However, Zubenelgenubi + KU~Lib will for ever have the honour of being the first to break the one-parsec-separation barrier.

\section{From February to October 2012: more cool wide binaries}

\begin{table}[] 
\caption{Wide multiple systems with cool dwarfs found in The Observatory series} 
\center
\begin{minipage}{1.0\textwidth}
\center
\begin{tabular}{l ll ccc} 
\hline\hline 
WDS	 	& Primary 				& Secondary		& ${\mathcal M}_{\rm total}$		& $s$ 			& $-U_g^*$		\\  
code		& 	 				& 	 			& [M$_\odot$]					& [kAU]			& [10$^{33}$\,J]	\\ [0.5ex]   
\hline 
KO 4		& NLTT 6496		 	& NLTT 6491 		& 0.34$^{+0.4}_{-0.5}$ 			& 5.7$^{+1.8}_{-1.2}$ & 7.7--8.2		\\ %
KO 5		& HD 212168 (A)		& DE2226--75 (C)	& 1.20$\pm$0.10				& 6.09$\pm$0.08	& 30$\pm$3		\\ %
KO 6		& LP 209--28		 	& LP 209--27		& 0.88:--1.04:					& 133--167		& 2.4--2.8			\\ %
FMR 83	& LSPM J0651+1843 	& LSPM J0651+1845 & 0.50$\pm$0.10				& 10$^{+6}_{-4}$	& 11$\pm$1		\\ [0.5ex]  %
\hline
\end{tabular} 
\end{minipage}
\label{T2} 
\end{table}

Koenigstuhl~1 and Zubenelgenubi + KU~Lib, with total masses and projected physical separations of 0.2\,M$_\odot$ and 1--2\,kAU, and 6--7\,M$_\odot$ and 200\,kAU, respectively, are the archetypes of the most fragile systems in their total-mass intervals.
Because of the large abundance of M dwarfs, it is natural to consider that there can exist a large number of systems with intermediate total masses and projected physical separations, i.e., 0.5--2.0\,M$_\odot$ and 10--100\,kAU.
Some of them were found by Gliese, Giclas or Luyten and have next gone unnoticed; others simply await discovery.
The main aim of the series ``Cool dwarfs in wide multiple systems'' at The Observatory is to find and characterise such intermediate-mass ultrafragile systems.

The four systems studied in The Observatory series are summarised in Table~\ref{T2}.
Koenig\-stuhl~4 (Caballero 2012) is a loosely-bound common-proper-motion pair of two bright mid-M dwarfs at about 19\,pc.
Koenigstuhl~5 (Caballero \& Montes 2012) is a triple system formed by a Sun-like {\em Hipparcos} star, a poorly investigated K dwarf, and an M8.5\,V-type star at 6.09\,kAU to the primary.
While these two systems were discovered serendipitously, Koenigstuhl~6 (Caballero et~al. 2012), an extremely fragile system of two late-K/early-M dwarfs, was found during the proper-motion survey for companions to Luyten stars (Section~\ref{section.200807}).
It was not only discovered by amateurs, but part of its photometric and astrometric characterisation was performed with an ``amateur'' telescope of 40\,cm.
Something similar happened with FMR~83, a common proper-motion pair of two identical mid-M dwarfs separated by about 10\,kAU, which was discovered by Rica (2012) and characterised by Rica \& Caballero (2012).

\section{October 2012: back to the future}

At the time of writing these lines, our near-future project is the ``Observatori Astron\`omic del Garraf Wide Pairs Survey''.
It involves over 20 Spanish double-star amateur astronomers supported by the Spanish Virtual Observatory.
It is a continuation of the Koenigstuhl and Luyten surveys, but in which the team uses the Aladin sky atlas for blinking two images of the Digital Sky Survey taken at very separated epochs (several decades) and for looking for common proper-motion pair candidates {\em by eye}.
The difference with the visual blinking method that was widely used in the middle of the 20th century is that we take advantage of virtual observatory tools and powerful computers that allow loading images and catalogues in seconds.
After surveying only 17\,\% of the sky ($\alpha$ = 0 to 12\,h, $\delta$ = --20 to +20\,deg), the team has discovered 1725 new pairs, which have been already tabulated by the Washington Double Star catalogue (T.~Tobal, priv.~comm. to the WDS) and will be presented in forthcoming publications (J.\,A.~Caballero, D.~Valls-Gabaud, E.~Solano et~al., in prep.).

The strengths of the wide pair survey are illustrated by GWP~1519, which is formed by G~45--26 (M1:\,V, $\sim$0.5\,M$_\odot$) and LSPM~J1101+0059 (M7:\,V, $\sim$0.1\,M$_\odot$).
At the most probable distance to the double, the angular separation of 41.3\,arcsec translates into a projected physical separation of $s \sim$ 2.4\,kAU, from where one deduces a low binding energy modulus of $|U_g^*| \sim$ --50\,10$^{33}$\,J.
Curiously enough, GWP~1519 was discovered during the very first days of the survey by N\'uria Miret, who at that time was a... High school student!



\small  
%
\section*{Acknowledgments}   
%
JAC is an {\em investigador Ram\'on y Cajal} of the CSIC at CAB.
Financial support was provided by the Spanish Ministerio de Ciencia e Innovaci\'on under grants 
AyA2011-30147-C03-02 and~-03. 

%

%
\end{document}